\documentclass[preprint,journal]{vgtc}       

\usepackage{mathptmx}
\usepackage{graphicx}
\usepackage{subfigure} 
\usepackage{times}
\usepackage{amsmath}
\usepackage{enumitem}
\usepackage[normalem]{ulem}
\usepackage[export]{adjustbox}

\usepackage[]{algorithm2e}
\usepackage{balance}

\usepackage{multirow}
\usepackage{stackengine}
\usepackage[bottom]{footmisc}

\newcommand\ClipImage[3]{
\settototalheight{\oH}{\includegraphics{#3}}%
\settowidth{\oW}{\includegraphics{#3}}%
\setlength{\rH}{\oH * \ratio{#1}{\oW}}%
\setlength{\rW}{\oW * \ratio{#2}{\oH}}%
\ifthenelse{\lengthtest{\rH < #2}}{%
    \setlength{\cW}{(\rW-#1)*\ratio{\oH}{#2}}%
    \adjincludegraphics[height=#2,clip,trim=0 0 \cW{} 0]{#3}%
}{%
    \setlength{\cH}{(\rH-#2)*\ratio{\oW}{#1}}%
    \adjincludegraphics[width=#1,clip,trim=0 \cH{} 0 0]{#3}%
}%
}

\newcommand{\revise}[2]{{#2}}

\usepackage[bookmarks,backref=true,linkcolor=black]{hyperref} 
\hypersetup{
  pdfauthor = {},
  pdftitle = {},
  pdfsubject = {},
  pdfkeywords = {},
  colorlinks=true,
  linkcolor= black,
  citecolor= black,
  pageanchor=true,
  urlcolor = black,
  plainpages = false,
  linktocpage
}

\onlineid{1284}

\vgtccategory{Research}
\papertype{Application/design study}

\title{\system: Connecting the Dots Between Features and Data to Explain Healthcare Models}

\author{Furui Cheng, Dongyu Liu, Fan Du, Yanna Lin, Alexandra Zytek, Haomin Li, \\ Huamin Qu, and Kalyan Veeramachaneni}
\authorfooter{
\begin{itemize}[noitemsep,nolistsep,leftmargin=0pt]
    \item Furui Cheng, Yanna Lin, and Huamin Qu are with Hong Kong University of Science and Technology. E-mail: \{fchengaa, ylindg, huamin\}@ust.hk.
    \item Dongyu Liu, Alexandra Zytek, and Kalyan Veeramachaneni are with Massachusetts Institute of Technology. Dongyu Liu is the corresponding author. E-mail: \{dongyu, zyteka, kalyanv\}@mit.edu.
    \item Fan Du is with Adobe Research. E-mail: fdu@adobe.com.
    \item Haomin Li is with the Children's Hospital of Zhejiang University School of Medicine. Email: hmli@zju.edu.cn.
\end{itemize}
}

\usepackage{xcolor}
\usepackage{tikz}
\usepackage{dsfont}
\usepackage{bm}

\newcommand{\etal}{\textit{et al.}}
\newcommand{\eg}{\textit{e.g.}}
\newcommand{\ie}{\textit{i.e.}}
\newcommand{\aka}{\textit{a.k.a.}}
\newcommand{\system}{VBridge}

\usepackage[most]{tcolorbox}
\definecolor{cback}{HTML}{EDEFF1}
\definecolor{cframe}{HTML}{B9C4CA}
\definecolor{cgrey}{HTML}{666666}
\newtcbox{\inlinebox}[1][]{enhanced,
 box align=base,
 nobeforeafter,
 colback=cback,
 colframe=cframe,
 size=small,
 left=0pt,
 right=0pt,
 boxsep=2pt,
 #1}
 
 \newtcbox{\inlinethinbox}[1][]{enhanced,
 box align=base,
 nobeforeafter,
 colback=cback,
 colframe=cframe,
 size=small,
 left=0pt,
 right=0pt,
 boxsep=1pt,
 #1}
 
\newtcbox{\nnboxx}[1][]{enhanced,
 box align=base,
 nobeforeafter,
 colback=cgrey,
 colframe=cgrey,
 colupper=white,
 size=small,
 left=1pt,
 right=1pt,
 boxsep=0.2mm,
 arc = 0mm,
 outer arc=0mm,
 #1}

\usepackage{tikz}
\usetikzlibrary{calc}
\makeatletter
\newcommand*\ccircle[2][draw]{\tikz[baseline=-0.65ex]
{\node[circle,#1,inner sep=.10ex] (char) {#2};}}
\makeatother
\newcommand{\nnbox}[1]{\ccircle[text=white,fill=cgrey]{#1}}

\definecolor{cback1}{HTML}{919191}
\definecolor{cframe1}{HTML}{ff0000}
\definecolor{darkblue}{HTML}{81ADD1}
\definecolor{lightblue}{HTML}{C7D8EB}
 \tcbset{
    box align=base,
    nobeforeafter,
    arc = 0mm,
    outer arc=0mm,
    colback=cback1, 
    colframe=cframe1,
    boxrule=0.5mm,
}

\newcommand{\cc}[1]{\textcircled{\small{#1}}}

\makeatletter
\newcommand{\myitem}[1]{%
\item[#1]\protected@edef\@currentlabel{#1}%
}
\makeatother

\abstract{
Machine learning (ML) is increasingly applied to Electronic Health Records (EHRs) to solve clinical prediction tasks.
Although many ML models perform promisingly, issues with model transparency and interpretability limit their adoption in clinical practice. Directly using existing explainable ML techniques in clinical settings can be challenging.
Through literature surveys and collaborations with six clinicians with an average of 17 years of clinical experience, we identified three key challenges, including clinicians' unfamiliarity with ML features, lack of contextual information, and the need for cohort-level evidence.
Following an iterative design process, we further designed and developed VBridge, a visual analytics tool that seamlessly incorporates ML explanations into clinicians' decision-making workflow.
The system includes a novel hierarchical display of contribution-based feature explanations and enriched interactions that \textit{connect the dots} between ML features, explanations, and data. 
We demonstrated the effectiveness of VBridge through two case studies and expert interviews with four clinicians, showing that visually associating model explanations with patients' situational records can help clinicians better interpret and use model predictions when making clinician decisions.
We further derived a list of design implications for developing future explainable ML tools to support clinical decision-making.

} 


\teaser{
 \centering
 \vspace{-0.5em}
 \includegraphics[width=16cm]{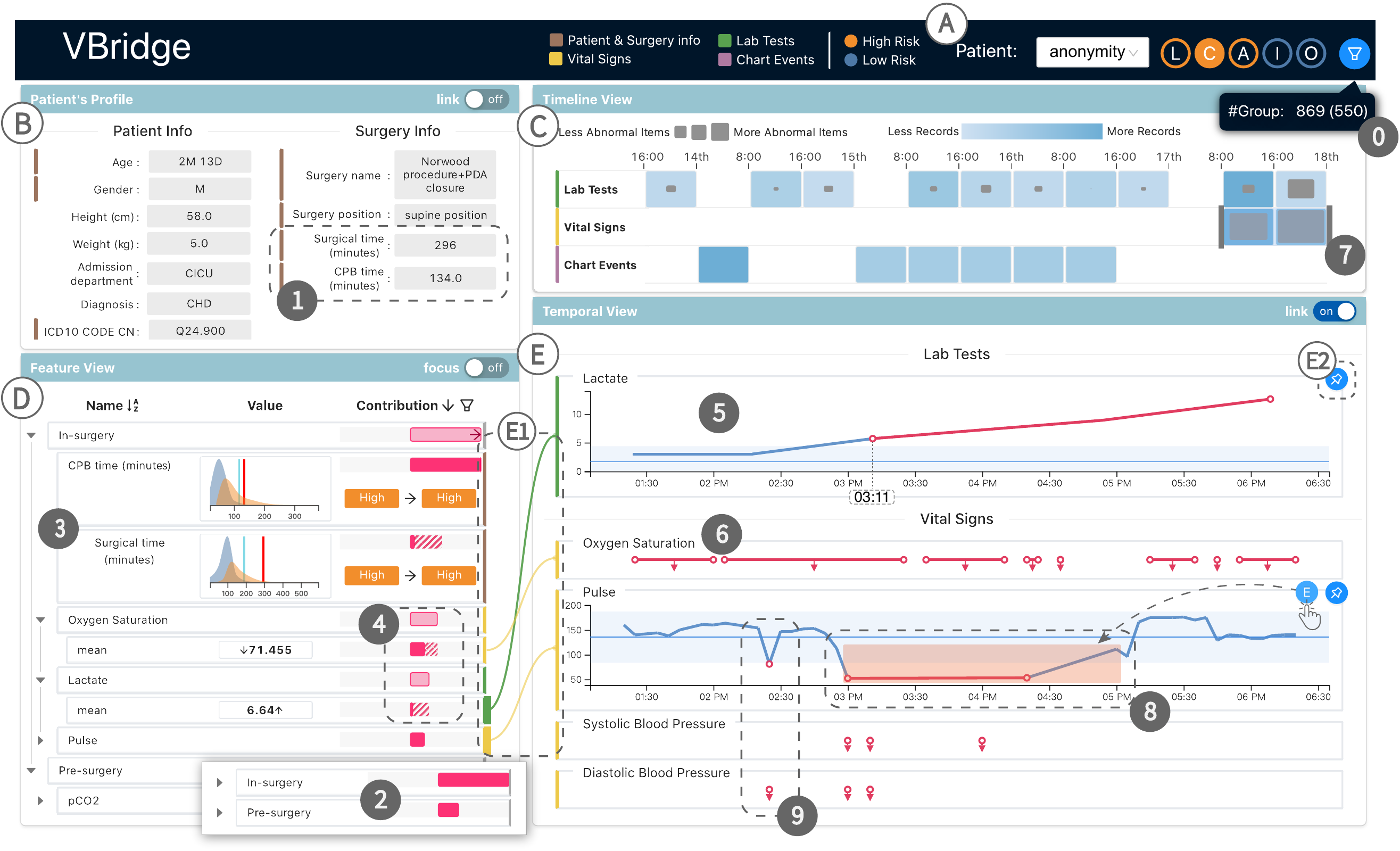}
  \vspace{-1em}
 \caption{The interface of \system~facilitates clinicians' understanding and interpretation of ML model predictions. 
 The header menu (A) allows clinicians to view prediction results, and to select a patient group for reference. 
 The profile view (B) and the timeline view (C) show a summary of the target patient's health records. 
 The feature view (D) shows feature-level explanations in a hierarchical display, linked to the temporal view (E) where healthcare time series are visualized to provide context for feature-level explanations. 
 \nnbox{0}-\nnbox{9} show the step by step progression of a use case in which a clinician used the system to explore model explanations. 
 After selecting a comparative group \nnbox{0} and viewing the patient's profile \nnbox{1}, he explored the feature-level explanations \nnbox{2}-\nnbox{4} to find the potential risk factors for the target patient. 
Then he referred to the patient's original records to gain an in-depth understanding \nnbox{5}-\nnbox{9}. 
 }
 \label{fig:teaser}
}

\begin{document}



\maketitle


\section{Introduction}
\label{sec:intro}

With the rapid proliferation of Electronic Health Records (EHRs), various prediction models based on machine learning (ML) techniques have been proposed for improving the quality of clinical care~\cite{rajkomar2019machine}.
An EHR stores an individual's health profile, from structured attributes like demographic information and medications, to unstructured ones, such as clinical notes and medical images.
Prediction models, trained on patients' EHR data, can be useful for a wide range of medical outcomes~\cite{harutyunyan2019multitask, rajkomar2018scalable}, including predicting a patient's \textit{remaining length of stay}, the likelihood of \textit{hospital readmission}, \mbox{and \textit{in-hospital mortality}.}

Despite efforts by researchers and developers to improve the performance of these prediction models, challenges remain -- including many associated with transparency and interpretability, which are particularly relevant in a highly regulated and risk-averse domain like healthcare~\cite{payrovnaziri2020explainable, sendak2020human}.
At the same time, XAI (eXplainable Artificial Intelligence) techniques and software tools continue to be developed, many of which have already proven powerful at elucidating the workings of ``black-box'' ML models.
Nevertheless, prediction models built upon modern ML techniques have not yet been widely and reliably used in clinical decision support workflows~\cite{jensen2012mining, miotto2016deep, rajkomar2019machine}.
By surveying related literature~\cite{ahmad2018interpretable, lipton2017doctor, payrovnaziri2020explainable, tonekaboni2019clinicians, wang2019designing, yang2019unremarkable} and working with 6 clinicians from a children's hospital, we found that the barriers preventing the application of XAI techniques in clinical settings are twofold.

First of all, clinicians engaging with XAI tools are often presumed to have sufficient technical expertise to understand and even improve ML models~\cite{ahmad2018interpretable}.
In reality, clinicians -- who may have little to no technical background -- are more likely to assess ML predictions through the lens of their domain expertise~\cite{sendak2020human} rather than understand and improve the ML model from the technical point of view.
This disconnection between the technology and its users is exacerbated by the fact that clinicians are rarely involved in discussions of explainability during the development of XAI tools~\cite{lipton2017doctor}.
As a result, the solutions provided by these tools are often intrinsically technical, leading to the difficulty for clinicians in understanding the ``explanations'' themselves~\cite{miller2019explanation, tonekaboni2019clinicians}.

In addition, clinicians' workflows are often guided by individual patients and may require tailored explanations based on each patient's EHRs (\ie, local explanations).
Among the copious XAI approaches that support local explanations, feature contribution is one of the most popular.
Approaches from this category illustrate the degree of contribution particular ML features make to a prediction outcome~\cite{payrovnaziri2020explainable}, which allows clinicians to directly compare model decisions with their own clinical judgment, especially when there is a disagreement.
However, although these approaches have been extensively studied within the XAI field, there are still several significant challenges preventing their actual use in healthcare. Clinicians working with ML may run into problems in the following areas:


\begin{itemize}[noitemsep,nolistsep]
\item \textbf{Understanding ML features.} 
Not every feature inputted to ML models is interpretable as-is by clinicians. For example, a patient's vital sign (\eg, in-surgery heart rate) will be transformed into multiple ML features, each represented by an aggregate value (e.g., SD (standard deviation) or Trend (linear slope)) within a period (\aka, feature engineering)~\cite{nemati2018interpretable}.
While easily understood by an ML model, this form of representation is almost certainly unfamiliar and non-intuitive to clinicians --
they may struggle to judge what, for example, a ``high'' Trend indicates, and the potential consequences.

\item \textbf{Connecting to patients' original records.}
Clinicians are more familiar with a given patient's original records than they are with ML features. In practice, they usually make decisions by referring to the raw data, such as laboratory test reports and vital signs from an anesthetic machine. However, feature contribution techniques only provide explanations on ML features, and do not deal with records directly~\cite{eck2017interpretation, ge2018interpretable, nemati2018interpretable, shrikumar2017learning}. How to seamlessly connect these explanations to patients' original records remains an open question, and one that is underexplored. 

\item \textbf{Aligning with evidence.}
Simply presenting a list of feature contributions in the form of numerical values does not allow clinicians to assess the trustworthiness of the model's predictions. 
Clinicians need to understand how feature contributions align with evidence-based medical practice~\cite{haynes1997evidence, tonekaboni2019clinicians}. \revise{}{In this research, we propose using cohort-level statistics, available through hospital records, to provide this evidence.} Clinicians can compare a target patient's feature values with reference values extracted from a cohort of similar patients. 
\end{itemize}

The aforementioned challenges motivate us to design and develop a visual analytics solution that can seamlessly integrate feature-level explanations into a clinician's decision-making workflow.
We followed a user-centric design process~\cite{wang2019designing,yang2019unremarkable} from the outset, working with 6 pediatric clinicians with an average of 17 years of work experience.
We derived design requirements from a pilot study with these clinicians; then, by observing their interactions with our early-staged system, we summarized two workflows -- forward analysis and backward analysis -- preferred by clinicians with different levels of expertise. 
These requirements and workflows guided the overall design and development of \system,
a \textbf{\underline{V}}isualization system that \textbf{\underline{Bridge}}s the gap between clinicians and ML models with tailored feature explanation algorithms and novel interaction and visualization techniques.

We adopted SHAP values~\cite{lundberg2018explainable} to generate contribution-based explanations of ML features and organized a large number of features in a hierarchy to facilitate interpretation.
We developed a novel visualization -- an interactive hierarchical feature list -- to present such explanations to clinicians in a user-friendly manner and integrated tailored visual designs to allow clinicians to conduct reference-value-based analysis and what-if analysis at the feature level.
To enable the connection between the feature explanations and the patient's raw records, we applied Deep Feature Synthesis~\cite{kanter2015deep} on EHR data to build traceable transformation paths between features and raw records.
Based on this, we present a tailored algorithm to identify the most influential records for a given feature.
The patient's original records are visualized in multiple coordinated views with different levels of detail. 
Various novel interactions, including linking and marking, help to visually associate the feature-level explanations and context information. 
The system was evaluated through two case studies and an expert interview with four clinicians, and results showed that our system is capable of supporting clinical decision-making.

To sum up, our contributions include:
\begin{itemize}[noitemsep,nolistsep]
    \item
    A summary of seven design requirements facilitating the interpretation of ML predictions to clinicians; 
    and the identification of two workflows describing how they work with ML models with feature-level explanations and needed context information. 
    
    \item
    A visual analytics system that integrates novel explanation algorithms and visualization and interaction techniques, to connect the dots between ML features, explanations, and health records for an improved clinicians’ decision-making workflow.\footnote{The code is available at https://github.com/sibyl-dev/VBridge}

    \item
    Two case studies and an expert interview demonstrating the usefulness and efficiency of our system.

\end{itemize}


\section{Related Work}


\subsection{Explainable Machine Learning in Clinical Predictions}

We categorize existing XAI techniques in clinical research based on whether the provided interpretability is intrinsic or post-hoc.


\textbf{Intrinsic interpretability.}  \label{sec:related:xai_intrinsic}
Models provide intrinsic interpretability by directly incorporating interpretability into their structures~\cite{choi2016retain, fleisher2007clinical, jalali2016interpretable, kho2012use, kwon2018retainvis, payrovnaziri2020explainable}.
Models in this category often use a simple structure to provide accurate and faithful explanations.
For example, Kho~\etal~\cite{kho2012use} used decision trees, which surface the set of rules driving the predictions, for predicting the genetic risk of type 2 diabetes.
Despite their intrinsic interpretability, the performance of these models is bounded compared to advanced ML models (\eg, deep neural networks), especially when handling complex clinical prediction tasks~\cite{harutyunyan2019multitask, xiao2018opportunities}.
Boosting and optimization techniques such as ensemble learning~\cite{jalali2016interpretable} can be used to enhance performance, but often at the cost of introducing additional complexity \mbox{that impairs interpretability}.

Recently, attention-based neural networks have begun to draw more focus ~\cite{payrovnaziri2020explainable}.
Such models do not directly inform clinicians of the reasons behind a prediction, instead highlighting the portion of historical data (\eg, clinical events) that have factored into it~\cite{choi2016retain, kwon2018retainvis}.
Although deep learning models can produce accurate predictions, attention-based explanations may cause information overload and confuse clinicians due to the lack of clarity around how prediction results relate to the areas of attention~\cite{payrovnaziri2020explainable}.
It is also challenging for attention-based deep learning models to support multimodal learning while preserving good interpretability~\cite{xiao2018opportunities}.

\textbf{Post-hoc interpretability}.  \label{sec:related:xai_posthoc}
Post-hoc methods take ``black-box'' ML models as inputs and then derive explanations for model predictions\revise{, keeping the underlying model performance intact}{}~\cite{che2016interpretable, cheng2020dece, NIPS2017_7062, ming2018rulematrix, ribeiro2016should}.
\revise{}{Unlike intrinsic interpretable models, post-hoc methods can be directly applied to existing models and thus are more flexible.}
One common approach is to use an intrinsically interpretable ML model to mimic a complex ``black-box'' ML model.
For example, Che~\etal~\cite{che2016interpretable} worked on acute lung injury (ALI) prediction and proposed a knowledge distillation method called mimic-learning, which uses gradient boosting trees to mimic the original deep learning model and provides rule explanations to clinicians.

Another type of work focuses on calculating feature contribution, which along with attention mechanism-based models, is considered to be one of the top popular approaches for supporting local explanations~\cite{payrovnaziri2020explainable}. For example, \revise{LIME -- a perturbation-based explanation approach introduced by Ribeiro \etal ~\cite{ribeiro2016should} -- has been leveraged to derive the contributions of features to the tasks of predicting girls' central precocious puberty~\cite{pan2019development} and autism spectrum disorder~\cite{ghafouri2019application}.}{}
Shapley Additive Explanations (SHAP)~\cite{NIPS2017_7062}, which build on the Shapley value from cooperative game theory~\cite{shapley1953value}, have been applied to explain hypoxemia predictions and support early prevention during surgery~\cite{lundberg2018explainable}.

\revise{In this work,}{Our work provides a post-hoc method for explaining existing models, in which} we adopt feature-contribution-based XAI approaches. In particular, we use SHAP to compute how each ML feature contributes to a particular prediction. 
We present a tailored visualization technique to display feature contributions to clinicians in a scalable and user-friendly manner. 


\subsection{Electronic Health Records Visualization}
We classify existing visualization techniques on EHR data~\cite{west2015innovative} based on the criterion proposed by Rind \etal ~\cite{Rind2013Survey} -- visualization for exploring health records from one patient or multiple patients.

\textbf{Individual patient records.} 
The goal of visualizing individual patient records is to provide individual patient summaries, as well as an efficient way to explore personal complex record data at different levels of detail.
A patient's clinical records contain longitudinal data representing patient visits over time.
One common way to summarize this history is through timeline-based visualizations, where events are placed on a horizontal timeline chronologically, using points or interval plots~\cite{plaisant2003lifelines, shahar1999intelligent, Shahar2006KNAVEII}.
\revise{Multiple stacked rows may be used when different event types are required to be distinguished~\cite{plaisant2003lifelines}.
To depict an individual event containing multiple attributes, one popular approach is to provide an interactive table that shows key clinical variables from one or multiple events over a period of time~\cite{bauer2006evaluating, Ghassemi2018ClinicalVis}.
Another common approach is using icons or glyphs to represent an event with multiple attributes~\cite{cao2011dicon}, where visual channels such as shape, size, and color are utilized for encoding different information.}{For events containing multiple attributes, glyphs~\cite{cao2011dicon} and additional tables~\cite{bauer2006evaluating, Ghassemi2018ClinicalVis} are used to visually summarize events and facilitate more detailed explorations.}
To further improve scalability, researchers have explored aggregation-based methods~\cite{krstajic2011cloudlines} and substitution-based approaches~\cite{Gotz2014DecisionFlow} to show frequent patterns instead of event details.
Another line of research focuses on event pattern searching, filtering, and grouping~\cite{wongsuphasawat2009finding, wongsuphasawat2012querying}, which supports fast and efficient data exploration.


In addition to discrete events, clinical signals collected during ICU or surgery are also commonly included in the EHR data. These are usually sampled at a higher frequency and can be viewed as continuous time series data.
Xu \etal~\cite{Xu2018ECGLens} used a spiral timeline to reveal periodic patterns of electrocardiogram data for arrhythmia detection.
\revise{In practice, to visualize EHR data from multiple sources, many visualization systems have adopted multiple coordinated views, where each view displays a certain type of information using a mix of the aforementioned techniques.
Our system used multiple coordinated views, a common approach in practice that leverages many of the advances offered by prior visualization techniques.}{Our system builds on advances offered by prior visualization techniques to visualize a patient's EHR data at different levels of detail.} We further tailored them for better interpreting ML predictions with feature-contribution-based XAI approaches.

\textbf{Multiple patient records.}
A number of scenarios require the analysis of multiple patient records, from patient cohort monitoring to observational clinical research.
\revise{A system that supports the visualization of multiple patient records must focus more on finding high-level patterns common to many patients.
Techniques such as aggregation~\cite{phan2007progressive, krstajic2011cloudlines}, small multiple~\cite{kwon2018retainvis, Jin2020CarePre} and flow-based representations~\cite{wongsuphasawat2011outflow, Gotz2014DecisionFlow} are often used to summarize a large set of EHR data.}{A large number of works focus on visualizing longitudinal EHR data~\cite{bernard2018using, caballero2017visual, Gotz2014DecisionFlow, gotz2014methodology, Jin2020CarePre, krstajic2011cloudlines, kwon2018retainvis, malik2015cohort, phan2007progressive, wongsuphasawat2011outflow}, where glyphs~\cite{bernard2018using, caballero2017visual} and flow-based representations~\cite{Gotz2014DecisionFlow, wongsuphasawat2011outflow} are often used for summarization. Other works focus on visualizing multivariate attributes or features transformed from the original records~\cite{alemzadeh2017subpopulation, bernard2015visual, Krause2014infuse, kwon2017clustervision, muller2020visual}. For example, Krause \etal~\cite{Krause2014infuse} designed a glyph to visualize the quality of a feature under different metrics.}
In this work, we used aggregation-based methods to extract reference values from a cohort of patients. We proposed small intense, simple, and embeddable visualizations to show the reference values. These are integrated with the feature explanation view and raw record data visualization view to allow reference-value-based analysis.

\section{Informing the Design}
In this section, we introduce the pilot study and detail the design requirements and analysis workflow distilled from the study.

\subsection{Pilot Study}
The pilot study allowed us to understand how clinicians expect to use ML prediction models with feature contribution explanations to support their clinical decision-making. We followed the design study methodology from Sedlmair \etal's work~\cite{sedlmair2012design} and designed the pilot study as follows.

\textbf{Participants:}
The study involved 6 clinicians (3 male and 3 female) from the Children’s Hospital of Zhejiang University School of Medicine (ZJUCH): 2 chief physicians from the Cardiac Intensive Care Unit (CICU) (\textbf{P1-2}) and 4 residents from the Cardiology Department (\textbf{P3-6}). 
Among them, \textbf{P1-3} are more senior, with an average of 24.5 years of work experience (20, 29, and 24 years respectively), while the others (\textbf{P4-6}) have an average of 10.5 years of experience (13, 10, and 8.5 years respectively).

\textbf{Presetting:}
The pilot study is based on a scenario of postoperative complication predictions. \revise{}{Patients may develop various complications after surgeries, some of which can be life-threatening. Predictions in an early phase can help clinicians identify high-risk patients and carefully choose postoperative caring plans.}
To support this scenario, we built a demo model on this prediction task.
We worked with a biomedical data scientist (DS) from ZJUCH -- a co-author of this paper, to carefully select a small set of features to train the model.
We use SHAP values to show features' contributions to the prediction result. 

\textbf{Process:} 
The study was divided into two sessions. 
We began the first session by performing one-on-one, semi-structured, hour-long interviews with all the participants. 
During the interview, the participants were presented with a low-fidelity mockup of our early system and taught some basic ML concepts. 
They were asked several questions about their understanding and concerns.
Based on the feedback collected from this session, we formulated the initial design requirements. Over the next three months, we developed a high-fidelity prototype system, holding weekly meetings with \textbf{DS} to make sure our implementation continued to meet requirements.

In the second session, we presented the prototype system to three participants (\textbf{P2}, \textbf{P3}, \textbf{P5}) separately. They were asked to explore the system freely and completed several prediction tasks then, during which they were encouraged to think aloud to explain their thoughts.
We observed, took notes, and collected their interaction processes.
We then held an open discussion with them to further understand their behavior.
The feedback collected from this round is further used to polish our design requirements and refine our system.


\subsection{Design Requirements}
\revise{Seven design requirements were ultimately summarized to guide the development of \system:}{We summarized seven design requirements and grouped them into: feature-level explorations (\inlinethinbox{Feature}), record-level explorations (\inlinethinbox{Data}), and explorations of feature-record connections (\inlinethinbox{Bridge}).}

\begin{enumerate}[nolistsep, label=\textbf{R\arabic*}]

    \item \label{r:feature-group}
    \inlinebox{Feature}
    \textbf{\revise{Group relevant features and show feature contributions hierarchically.}{Show features in a hierarchical structure.}} 
    All participants (\textbf{P1-6}) confirmed that it is challenging to explore hundreds of features extracted from diverse and heterogeneous sources.
    They all agreed with the idea of grouping relevant features semantically for a better exploration experience. For example, the aggregation values (\eg, Mean, SD, and Trend) computed from the same series of data (\eg, pulse) can be reasonably grouped.
    
    
    \item \label{r:feature-reference}
    \inlinebox{Feature}
    \revise{}{\textbf{Provide features' reference values}.
    All participants (\textbf{P1-6}) agreed that the features, especially aggregate values that are unfamiliar to clinicians (\eg, SD and Trend), should be presented alongside reference values, which describe the range of values that are considered normal. 
    Because there are no existing reference values for most of the features, the system should calculate them using data from a relevant cohort.}
    
    \item \label{r:feature-explore}
    \inlinebox{Feature}
    \revise{\textbf{Offer users controllability when exploring feature contribution results.}}{\textbf{Provide flexible interactions} to support on-demand explorations.}
    Participants follow different strategies when exploring features. 
    Some participants (\textbf{P1}, \textbf{P3-5}) are only interested in the riskiest factors (\ie, features with high positive contributions), while some (\textbf{P2, P6}) were also interested in negatively contributing features that could be helpful for lowering the surgery risks in the future.
    Thus, the system should enable \textbf{sorting} and \textbf{filtering} to support different exploration paths.
    \revise{}{In addition, \textbf{P1} and \textbf{P2} expressed a further need to conduct what-if analyses on abnormal features to better understand their effects on predictions.}

    \item \label{r:record-summary}
    \inlinebox{Data}
    \textbf{\revise{Visually summarize the patient's records from diverse heterogeneous sources}{Provide an overview of the patient's records}.}
    Patients have complex medical records, especially ICU patients, who may have substantially more data available than general patients. Good visualizations summarizing a patient's visiting history 
    \textit{``can save us a great amount of time in [familiarizing ourselves with] a patient's background''}, \textbf{P2} and \textbf{P6} confirmed.
    
    \item \label{r:record-reference}
    \inlinebox{Data}
    \textbf{Show record details with reference values.} 
    Similar to \ref{r:feature-reference}, participants (\textbf{P1}, \textbf{P3-5}) suggested that showing patients' historical health record details along with reference values helps them to make more informed decisions. \textbf{P1} would like to know whether these values are within the 95\% confidence interval (CI) of a statistical summary of similar patients.
    
    \item \label{r:bridge-association}
    \inlinebox{Bridge}
    \textbf{Visually associate feature(s) with the patient's records.}
    All participants (\textbf{P1-6}) expressed the need to check the original records (\ie, medical events) of particular features that interested them. \textit{``Manually checking without any clues would take me 10-20 minutes''}, \textbf{P5} commented.
    Thus, visual associations of such correlations along with a tailored interaction mechanism should be enabled to support efficient back-and-forth analysis between features and their relevant original records.
    
    \item \label{r:bridge-highlight}
    \inlinebox{Bridge}
    \textbf{Highlight temporal value patterns} that are influential to feature(s). 
    Three participants (\textbf{P1}, \textbf{P3}, \textbf{P5}) expected the system to highlight high-risk time periods from long-lasting vital sign records related to the feature under investigation. \textbf{P1} showed particular interest in influential time periods containing a series of data points, rather than isolated anomalous data points which could be caused by errors.
    
\end{enumerate}

\subsection{Analysis Workflow}
\label{sec:workflow}
After analyzing the user interaction patterns and discussion notes from the second pilot study session, we summarized two general analysis workflows: \textbf{forward analysis} and \textbf{backward analysis}. 

\begin{enumerate}[nolistsep]
    \setlength\itemsep{-0.1em}
    \myitem{\textbf{WF}} \label{f:forward} 
    \textbf{Forward analysis:}
    clinicians \revise{}{inspect the data in an order similar to that of the direction of the data processing flow (\ie, original records $\rightarrow$ features $\rightarrow$ predictions). They first} make their own prediction \revise{using evidence collected from the original records, and then compare their predictions with the model predictions and explanations.}{based on the patients' original records, then compare their predictions with the model predictions, and finally check explanations to make decisions.}
    Senior clinicians such as \textbf{P1} and \textbf{P2} preferred to start by viewing the patient's profile and forming initial hypotheses.
    They would then look directly at the original records (\ref{r:record-summary}, \ref{r:record-reference}) and check potentially influential observational records (\eg, in-surgery lactate records).
    After making their own predictions based on this evidence, they seek confirmation from ML models and feature explanations (\textbf{R1-3}).
    If the model prediction and explanations agreed with their expectations --  assigning high contribution values to the factors they thought were risky -- their trust in the model was enhanced.
    If it didn't, they would refer to the patient's original records relevant to the features they were investigating to find more evidence-based on reference values (\textbf{R5-7}).
    They would either reject the model prediction, or would gain new knowledge based on this evidence.

    \myitem{\textbf{WB}} \label{f:backward} 
    \textbf{Backward analysis:}
    clinicians \revise{}{inspect the data in the opposite direction as the data processing flow. They} check the model predictions and explanations first, and then trace back to the original records, finding evidence to \revise{confirm their hypotheses}{support their decisions}.
    When clinicians started without a clear diagnostic prediction, they began with a feature list with contribution explanations (\textbf{R1-3}). They then identified a set of features for further investigation.
    For static and familiar dynamic features (\eg, in-surgery pulse), they compared the feature contributions with their expectations.
    For unfamiliar features, like SD or Trend of in-surgery systolic blood pressure, they preferred to check the details in the original records (\textbf{R5-7}).
    Sometimes these records were not sufficient to verify their hypothesis. They would then check the summary information of the original records (\textbf{R4}) in order to obtain different records from the same time, for correlation analysis. 
    
    
\end{enumerate}


\section{Predictive Modeling}\label{sec:modeling}


In this section, we first introduce the dataset, along with the prediction task we use as a running example for our research. Then we introduce how we extract ML features and generate explanations.

\subsection{Data}
\revise{Our system uses the Paediatric Intensive Care (PIC) Database\footnote{PIC dataset: \url{http://pic.nbscn.org/}}}{\system{} takes structured EHR data collections as input. These are often organized as relational databases. In this work, we use the Paediatric Intensive Care (PIC) Database~\cite{zeng2020pic} as an example}, which contains de-identified clinical data of paediatric patients admitted to ZJUCH.
In particular, the dataset collects over $14,000$ hospital admissions from $12,000$ unique paediatric patients, aged 0–18 years, admitted to the critical care unit between 2010 and 2019.
The PIC follows the same paradigm to store ICU patient clinical records as the widely-studied Medical Information Mart for Intensive Care (MIMIC-III) dataset~\cite{johnson2016mimic}, but puts more emphasis on paediatric patients.
The dataset encompasses a number of information types, including demographics, surgery information, high-resolution vital sign measurements during surgery, laboratory test results, symptoms, medications, diagnostic codes, and mortality.

\subsection{Running Example: Surgical Complication Prediction}
\label{sec:modeling:running_example}
To concretize our system's contributions, we utilize a running example -- predicting complications after cardiac surgery -- from a case study involving two clinicians from the ZJUCH team (\textbf{P1}, \textbf{P5}).
This team was interested in using ML models to predict whether a patient is at risk of developing five types of complications after cardiac surgery: lung, cardiac, arrhythmia, infectious, and others, which are each annotated with their first letter (\ie, L, C, A, I, O). A patient may experience multiple postoperative complications.

Working with the team and starting with the entire PIC dataset, we first selected $1,826$ patients who underwent cardiopulmonary bypass-supported cardiac surgery. 456 (25.0\%) of these patients developed postoperative complications.
From the medical records of these patients, we mainly extracted three types of static features (demographics, surgery information, and diagnosis results) and three types of dynamic features whose values change over time (lab tests, surgery vital signs, and chart events\footnote{Chart events contain patients' routine vital signs (not during surgery) and additional information like inputs and outputs.}).
In total, there were 1,724,805 lab test events, 450,989 chart events, and 754,213 data points from vital signs.
In this example, our goal was to build 5 individual binary classifiers, each predicting one of the five complication types.

To be accepted by an ML model, a patient's raw medical data must be transformed into an ML-understandable format (\aka, feature engineering) -- namely, a feature vector (Fig.~\ref{fig:featuretools}\cc{C}).
Multiple feature vectors compose a feature matrix with each row describing one patient. 
Given the feature matrix and the target prediction column, in order to obtain the best ML model for our task, we applied Cardea~\cite{alnegheimish2020cardea} -- an automated machine learning (AutoML) framework for EHR data. The framework evaluated 8 classifiers
whose hyperparameters were optimized using AutoML for a higher performance score -- the averaged AUC of $10$ cross-validation folds \revise{}{(see Appendix)}. 
Finally, we obtained five models, each of which performed the best for one complication.

\subsection{Feature Extraction}
\label{sec:modeling:feature_abstraction}

\begin{figure}[!t]
    \centering
    \includegraphics[width=1\linewidth]{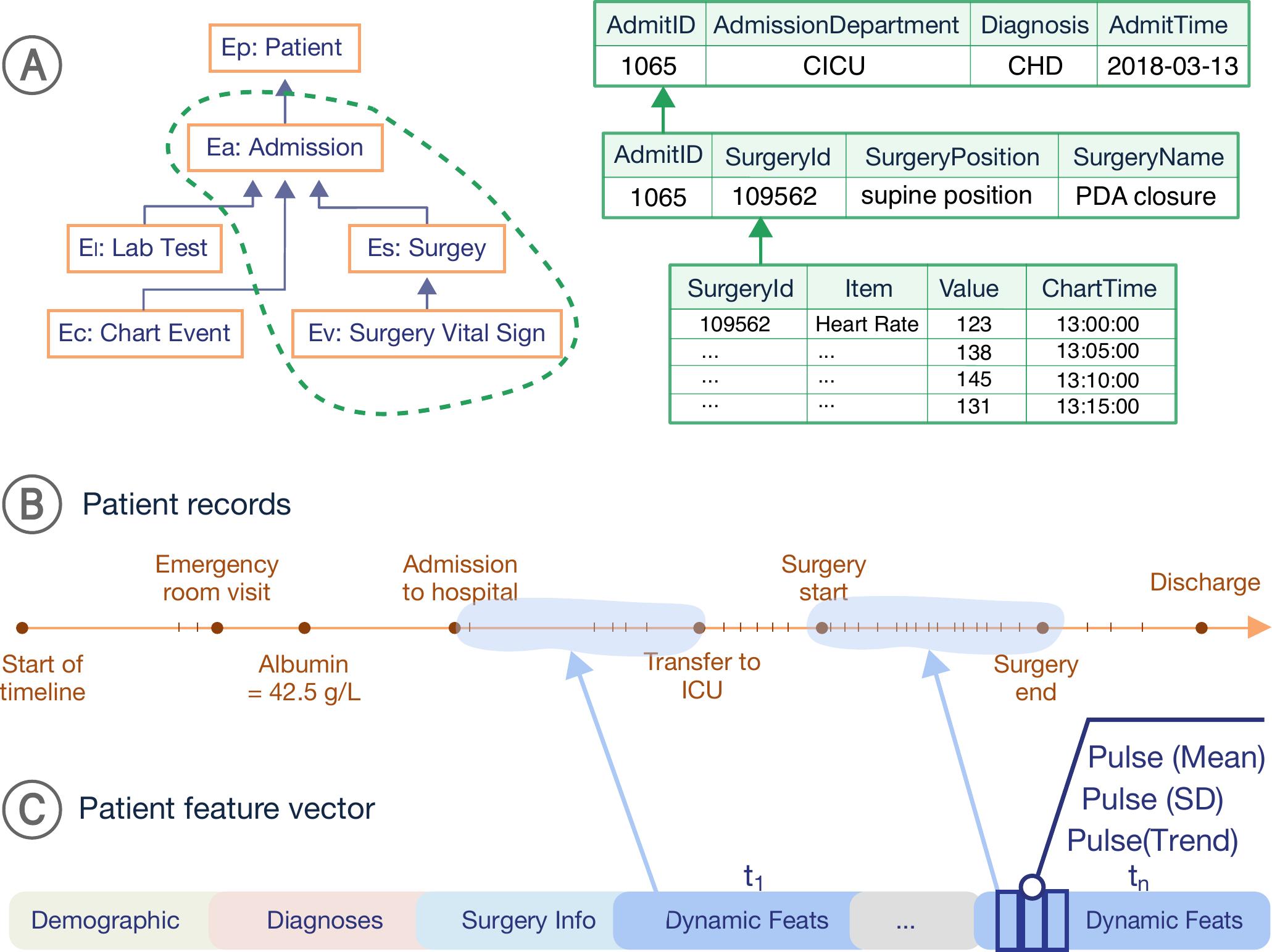}
    \caption{Multiple-sourced EHR data is saved in different connected tables or entities (A). We use a DFS algorithm to extract it as a set of chronologically ordered patient
    records (B) and an ML feature vector (C). }
    \label{fig:featuretools}
\end{figure}


As shown in Fig. \ref{fig:featuretools}\cc{A}-left, EHR data from different sources is described as different entities, such as Admission ($E_a$), Lab Test ($E_l$), and Surgery ($E_s$).
Entities are connected by reference keys (Fig. \ref{fig:featuretools}\cc{A}-right).
In our running example of surgical complication prediction, we worked with our clinician collaborators and identified six feature types, both static and dynamic, with which to compose a patient's feature vector. Our target patients are those who underwent cardiac surgery.
To that end, we chose Surgery ($E_s$) as the target entity, and extracted the associated features including patient profile from $E_p$, surgery information from $E_s$, low-resolution time series (lab test and chart events) from $E_l$ and $E_c$, and high-resolution time series from $E_v$.

\textbf{Assembling patient feature vectors with DFS.}
We adapted Deep Feature Synthesis (DFS)~\cite{kanter2015deep} -- an algorithm that automatically generates features out of relational tables -- for our scenario to ensure that the connection between a patient's feature vector and raw records is traceable (\ref{r:bridge-association}).
It works by following relationships between tables to a base field (\eg, SurgeryId) and then sequentially applying transformation functions along the path to create the final feature. 
In the end, the algorithm will recursively extract all the associated features to our target entity (Surgery $E_s$) and each feature corresponds to a traceable path of length $l$ between the final feature value and the raw record(s) of the source entity.
The path is important for the purpose of visualization and the identification of influential records (\ref{r:bridge-association}, \ref{r:bridge-highlight}).

\subsection{Feature Explanation}
\label{sec:modeling:feature_explanation}

\revise{We applied feature-contribution-based XAI approaches to explain the outputs of ML models. In particular, we used SHAP values~\cite{NIPS2017_7062} -- a game theoretical approach that connects optimal credit allocation with local explanations using the Shapley values -- to compute how each feature contributes to a particular prediction. A number is associated with every feature, where the number's value suggests the level of contribution and its sign indicates a positive (increase risk) or negative (decrease risk) contribution. 
However, providing only feature-level explanations is not enough for clinicians, who may want to check the patient's original medical records (\ref{r:record-summary}, \ref{r:record-reference}) and thus wish to know which records (particularly for dynamic features, Fig.~\ref{fig:featuretools}\cc{C}) are responsible for the feature of interest (\ref{r:bridge-association}, \ref{r:bridge-highlight}).}{We applied SHAP values to provide feature-level explanations. However, for features that are unfamiliar to clinicians (\eg, Trends), such explanations are not sufficient. Clinicians wish to further understand which time periods within the records (Fig.~\ref{fig:featuretools}\cc{C}) are responsible for the feature of interest (\ref{r:bridge-association}, \ref{r:bridge-highlight}).} 

A common approach is occlusion sensitivity~\cite{Zeiler_Fergus_2014}.
However, simply removing several medical records and observing how the prediction changes is not a feasible solution, because a surgical patient usually produces thousands of records -- meaning that the model will not be sensitive if only a small number of records are removed\revise{ and uncertainty will be increased}{}. \revise{}{A similar approach, observing how relevant features change under the occlusion, has the same sensitivity issues.}
\revise{Here we propose a novel solution that still leverages an occlusion-based method but calculates the influence to the feature value rather than the prediction result. We argue that the benefits are twofold: (1) this solves the underlying sensitivity issues; (2) clinicians can more easily interpret such influence using their domain knowledge, especially when the reference value is provided. 
Consider a scenario where a patient's $mean(pulse)$ has a largely positive contribution to the prediction, and this feature value is significantly higher than the reference value computed from the control group. If the pulse records of this patient over a certain short time period largely increase the value of the $mean(pulse)$ feature, clinicians can confirm that this time period is critical and investigate it. }{To solve the underlying sensitivity issues, we first calculate the influence of the records on the \textbf{relevant features' values} and identify the most influential time periods, using occlusion-based methods introduced below. Then we filter the influential record segments that push the relevant features' value away from the average level (\ie, the reference value).
Consider a scenario in which a patient's $Pulse(Trend)$, a major contributing feature, is significantly higher than the reference value. Clinicians may want to know during which period the records specifically cause a sudden increase in feature values, rather than all influential periods. }

\textbf{Computing record influence on a dynamic feature}. 
Given a window size of $k$, a series of temporally ordered records $E = [x_1, x_2, ..., x_T]$, and a result array $\bm{v}$ of length $T$ initialized to all 0s, we iteratively replace -- or ``occlude'' -- segments $E_{t:t+k}$ with some set of values, and increment $t$ by after each step.
We use window size $k$ to reduce the impact of data quality issues and focus more on a segment of data (\ref{r:bridge-highlight}). 
We propose the use of a linear curve fit to the points in the window, which maintains smoothness while removing unique features within the window.

After each occlusion step, we recalculate the feature value, and store the change between the original feature value $x$ and the updated feature value $x'$ in the corresponding indices of $\bm{v}$: $\bm{v}_{t:t+k} = \bm{v}_{t:t+k} + (x - x')/\text{abs}(x)$.
The results in $\bm{v}$ show the relative total influence of each record in $E$, based on how much and in which direction the feature changes when this point is removed. Notably, the real time computation of $x'$ is possible because we store the traceable path between the relevant raw records and the feature value (Sec.~\ref{sec:modeling:feature_abstraction}).


\textbf{Identifying the most influential time periods}.
Now that we have obtained an array of influence values $\bm{v}$, the next step is to highlight the most influential time periods (\ref{r:bridge-highlight}).
This involves finding a threshold $\theta$ and identifying a list of segments $\bm{V}_{seq}$ with values above that threshold. 
Given that parametric approaches such as use of a Gaussian tail can be flawed when parametric assumptions are violated (\eg, that the data follow Gaussian distributions), we adopted a non-parametric method without statistical assumptions.
This method is adapted from the dynamic threshold computing method proposed by Hundman \etal~\cite{hundman2018detecting}.
We pick a threshold from the set: $\bm{\theta} = \mu(\bm{v}) + z\sigma(\bm{v})$, where $z \in \bm{z}$ is an ordered set of positive values indicating the number of SD ($\sigma$) above the mean ($\mu$). The optimal $\theta$ is determined by:
$$\underset{\theta}{argmax} \quad \frac{\Delta\mu/\mu(\bm{v}) + \Delta\sigma/\sigma(\bm{v})}{|\bm{v}_a| + |\bm{V}_{seq}|^2}	$$
where 
$\bm{v}_a = \{v_i \in \bm{v} | v_i > \theta\}$,
$\Delta\mu = \mu(\bm{v})-\mu(\bm{v} \setminus \bm{v}_a)$,
$\Delta\sigma = \sigma(\bm{v})-\sigma(\bm{v} \setminus \bm{v}_a)$, and
$\bm{V}_{seq} = \text{continuous sequences of } v_a \in \bm{v}_a$
The goal is to find a threshold $\theta$ that -- once all values above it are eliminated -- would lead to the maximum percent decrease in mean and SD of $\bm{v}$. Then we can obtain $\bm{V}_{seq}$ which represents the set of most ``exceptionally'' influential segments for a feature. We propose a novel visualization for showing this information to clinicians, detailed in Sec.~\ref{sec:view_temporal}.


\section{\system}

In this section, we continue with our running example -- surgical complication prediction -- to introduce our system designs.

\begin{figure}[!htb]
    \centering
    \includegraphics[width=1.0  \linewidth]{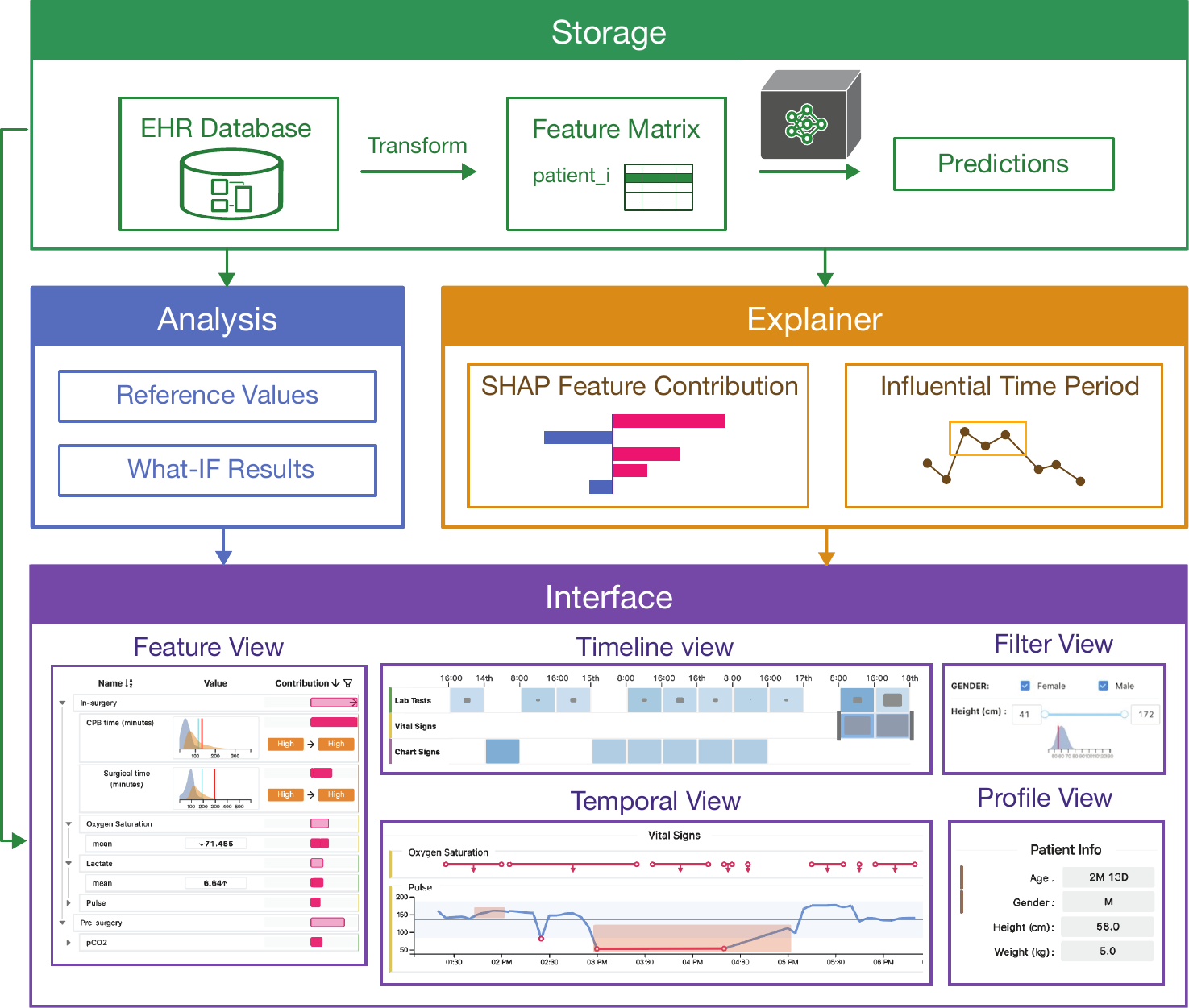}
    \caption{The system architecture of \system, which consists of four tightly connected modules: a \textcolor[HTML]{1f9059}{\textbf{storage}} module, an \textcolor[HTML]{5574b8}{\textbf{analysis}} module, an  \textcolor[HTML]{da9227}{\textbf{explainer}} module, and an \textcolor[HTML]{7b58a5}{\textbf{interface}} module.}
    \label{fig:system_arc}
\end{figure}

\subsection{System Overview}
\label{sec:system}

Fig.~\ref{fig:system_arc} illustrates the system architecture and the interactive analysis pipeline it supports.
\system~ comprises four major modules: (1) storage, (2) analysis, (3) explainer, and (4) interface.
The storage module saves all original patient records, a feature matrix with each row representing a patient's clinical features, and the ML prediction results.
The analysis module supports dynamic calculation of reference values when a cohort of patients is selected, and real-time computation of the results for what-if analysis.
The explainer module uses SHAP values to represent feature contributions, and identifies influential time periods for a given feature.  
Lastly, the interface module supplies multiple visual views, allowing a clinician to carry out his/her analysis using either a forward or backward workflow.

To show how the five views in the interface are connected, we assume that a clinician is using the backward analysis workflow to investigate the risk that a patient will have postoperative complications.
First, he picks the patient and complication of interest from the top menu (Fig. \ref{fig:teaser}\cc{A}). The five icons beside the selection box show the prediction results for the five types of complications -- orange for positive and blue for negative.
Next, he views the patient demographic, surgery, and admission information through the \textit{Profile View}, identifies a cohort of patients as the reference patient group using the \textit{Filter View}\revise{ to facilitate evidence-based analysis. He can observe the patient number on the top menu \mbox{as well (Fig. \ref{fig:teaser}\nnbox{0}).}}{, and checks the patient number on the top menu \mbox{(Fig. \ref{fig:teaser}\nnbox{0}).}}

Next, he begins formally investigating from the \textit{Feature View}, which hierarchically shows ML-features, their contributions to the prediction result, and their reference values (\ref{r:feature-group}, \ref{r:feature-reference}, \ref{r:feature-explore}).
To further investigate the features of interest, he should check the original records of these features using the \textit{Temporal View}; this view visualizes how a patient's clinical records change over time, along with the calculated reference range (\ref{r:record-reference}) and the influential periods (\ref{r:bridge-highlight}).
Multiple types of visual association, such as linking, filtering, and highlighting, are enabled to support the back-and-forth analysis between the \textit{Feature View} and the \textit{Temporal View} (\ref{r:bridge-association}).
If the original records of the target feature (\eg, Heart Rate) are not sufficient to verify the hypothesis, the clinician should then refer to the \textit{Timeline View} to understand the overall situation and select contemporary medical records from other relevant features for further investigation (\ref{r:record-summary}). 

\subsection{Feature View}
\label{sec:view_feature}

The \textit{Feature View} (Fig. \ref{fig:teaser}\cc{D}) aims to allow clinicians to explore and understand the model's behavior at the feature level. To provide more consistent representations of these massive features, we have grouped relevant features according to suggestions from our clinician collaborators (\ref{r:feature-group}). We first group the features (\eg, Mean, SD, and Trend) that were extracted from the same series of medical events (\eg, Pulse Records). We further divide the features (or groups) \revise{based on when the information is obtainable}{according to their temporal occurrence}, which includes ``pre-surgery'' features (\eg, demographics) and ``in-surgery'' features (\eg, vital signs). 

\begin{figure}[!htb]
    \centering
    \includegraphics[width=1.0\linewidth]{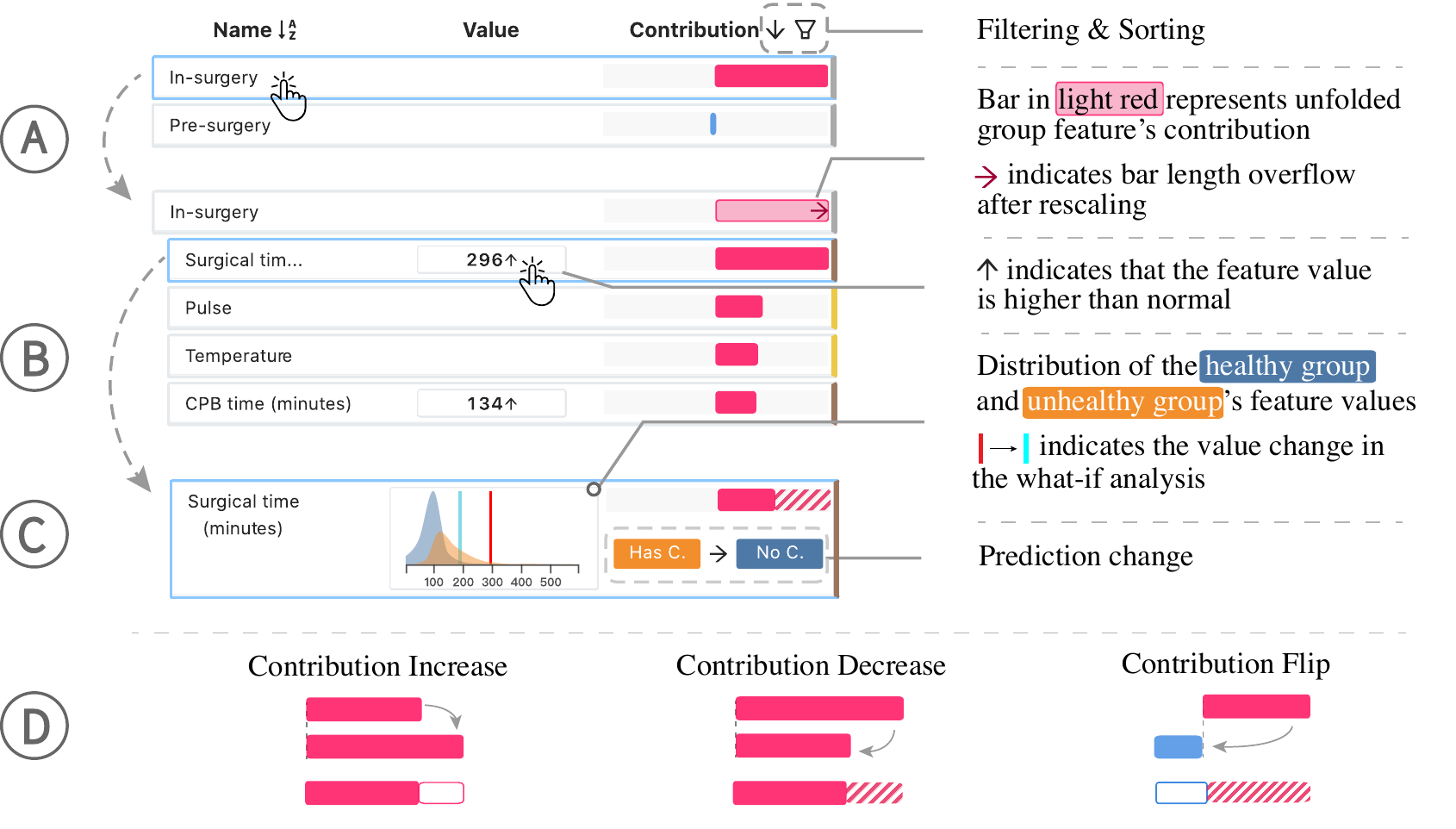}
    \caption{Feature View. The design of the hierarchical feature list with different detail levels (A-C) and the design for visualizing contribution changes in the what-if analysis (D).}
    \label{fig:feature-view}
\end{figure}

\textbf{Hierarchical display (\ref{r:feature-group}).}
We visualize the features in a hierarchical list where each row represents a feature or a feature group (Fig. \ref{fig:feature-view}). For each feature, we present the value and its contribution to the model prediction. 
\revise{We use the SHAP algorithm (Sec.~\ref{sec:modeling:feature_explanation}) to calculate these contributions and deploy a horizontal bar to visually encode the contribution value}{We visually encode the contribution value with a horizontal bar}, where the color encodes its sign (red for increasing complication risks and blue for decreasing risks) and the length encodes its magnitude (Fig. \ref{fig:feature-view}\cc{A}). 
For a feature group, we calculate group-level contributions by summing up the included features' additive contributions. 

Based on the definition of Shapley values~\cite{shapley1953value}, group-level contributions can be explained as an approximation of the effects of removing this group of features from the model. 
Because clinicians have different goals or levels of knowledge, some expect to investigate the most fine-grained level of features (\eg\, SD and Trend) while others may stop at the group level.
The hierarchical feature list matches their demands well in this regard.
Sorting and filtering by contributions are also supported to offer clinicians more control during \mbox{explorations (\ref{r:feature-group}).}

\textbf{References from cohorts (\ref{r:feature-reference}).} 
\revise{Reference values help clinicians better understand what feature values actually indicate. In our system, we calculate reference values from a group of similar patients selected by clinicians.
Selected patients are divided into a low-risk group (\ie\ no complications) and a high-risk group (\ie\ at least one complication). 
We use a 95\% CI (confidence interval) -- a common threshold used in clinical research~\cite{altman2001revised} -- of the low-risk group's mean value of a feature (\ie\ estimated as $Mean\pm 1.96\times SD$) as the reference range.}{In \system, reference values are calculated from a relevant cohort (\eg, patients in the same age range) selected by users through the \textit{Filter View}. The selected cohort is further divided into a low-risk group (\ie\ no complications) and a high-risk group (\ie\ one or more complications). We use the 95\% CI of the low-risk group's mean value as the reference value range.}
We use an upward/downward arrow to indicate whether a value is beyond the upper/lower bound of the reference range (Fig. \ref{fig:feature-view}\cc{B}).

Clinicians can click the value area to inspect detailed value distributions of the low-risk group and the high-risk group (Fig. \ref{fig:feature-view}\cc{C}).
For a continuous feature, the distribution is visualized with area charts, where a red line indicates the position of the feature value in relation to the target patient. 
For a categorical feature, we use bar charts to depict the distribution rather than area charts.

\textbf{What-if analysis (\ref{r:feature-explore}).}
In evidence-based clinical practice, clinicians pay a lot of attention to anomalous records (\eg, low Oxygen Saturation Rate) in the process of clinical reasoning.
Our system marks values out of the reference range as anomalies. Clinicians are particularly interested in highly contributed features with anomalous values. 
For example in Fig. \ref{fig:feature-view}\cc{B}, the surgery time (296 minutes) is noted as an exceptionally high value by the upper arrow, and it also makes the highest contribution to the prediction.
Our clinician collaborators had expressed strong interest in such cases, leading us to ask: 
If such a value is normal, does it still make \mbox{a large contribution?}

To answer this question, we designed a reference-value-based what-if analysis technique. Unlike open-ended what-if analysis techniques~\cite{wexler2019if}, we focus on one abnormal feature at a time, and make a minimal change to fit the reference range\revise{, setting it to the upper or lower bound depending on whether it is higher or lower than the references}{\ (\eg, setting a high blood-pressure-related feature value to the upper bound of its reference range)}. \revise{}{Then we calculate and visualize changes in the prediction result and the target feature's contribution (Fig. \ref{fig:feature-view}\cc{C} - bottom right).}
\revise{}{We designed visualizations to encode the contribution changes while reserving the original contribution (\ie, solid and dashed area) as context (Fig. \ref{fig:feature-view}\cc{D}).}
This approach provides clinicians with the most efficient and familiar way to verify their findings, especially when they aren't well-practiced in setting feature values.

\subsection{Temporal View}
\label{sec:view_temporal}

\begin{figure}[!htb]
    \centering
    \includegraphics[width=1.0\linewidth]{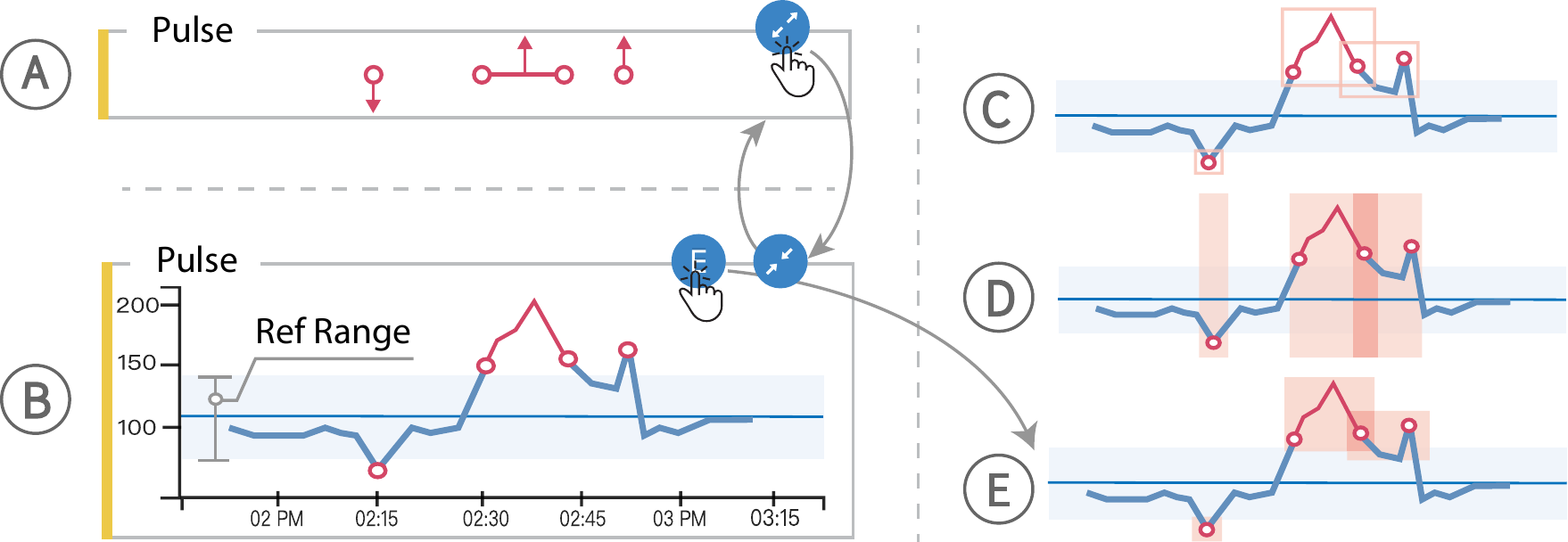}
    \vspace{-1em}
    \caption{Temporal View. The collapsed version (A) and the expanded version (B) of our time series record visualization, with anomalies highlighted.  (C-E) show three design alternatives for highlighting influential time periods, where (E) is our last choice.}
    \label{fig:alternative}
    \vspace{-1.5em}
\end{figure}

The \textit{Temporal View} visualizes a list of time series -- each representing a type of time-varying clinical feature (\eg., Heart Rate) -- in order to provide context for feature-level explanations (Fig. \ref{fig:teaser}\cc{E}).
When clinicians find interesting features in the feature view (\eg, Mean of Oxygen Saturation), they can append the corresponding time series records to the temporal view for further inspection (\ref{r:record-reference}).

Each time series is visualized as a line chart (Fig. \ref{fig:alternative}\cc{B}). We use the translucent blue area with a horizontal line in the middle to show the reference range (\ie, 95\% CI\revise{ as introduced in Sec.~\ref{sec:view_feature}}{}) and the mean value from the selected patient group.
This design is familiar to clinicians and has frequently been used in clinical research~\cite{2012SimTwentyFive}.
\revise{}{In this paper's running example, children's observed values (\eg, Pulse) vary significantly in different situations. In response, we compute reference values dynamically according to the selected group of patients similar to the feature reference values (Sec.~\ref{sec:view_feature}).}
We further empower the design to support the analysis of anomalies, concurrent patterns, and influential segments. 

\textbf{Highlighting anomalous records and enabling concurrent patterns analysis.}
Inside the line chart (Fig. \ref{fig:alternative}\cc{B}), we use red dots and line segments to highlight out-of-reference-range records and time periods. 
To support clinicians inspecting multiple time series at the same time for concurrent pattern analysis (Fig. \ref{fig:teaser}\nnbox{9}), we use a space-intensive design to only show the out-of-reference-range segments (Fig. \ref{fig:alternative}\cc{A}). The arrow direction indicates whether a point (segment) is above or below the reference value range, whose design is consistent with the one used in the \textit{Feature View} (Fig. \ref{fig:feature-view}\cc{B}).

\textbf{Highlighting influential value patterns.}
Reference values provide clinicians with an evidence-based method for insight verification. However, clinicians are also curious to see how a ML model judges the influence of certain time periods, such as high-risk periods captured by the model (\ref{r:bridge-highlight}).
We use the algorithm described in Sec.~\ref{sec:modeling:feature_explanation} to identify the most influential \textbf{non-overlapping} time segments $\bm{V}_{seq}$ and highlight them in the line chart (Fig. \ref{fig:alternative}\cc{E}).
However, multiple features (Mean, SD, Trend, etc.) may be associated with the same series of records (e.g, Pulse), so segments that are influential to different features can overlap.
These overlapped areas often suggest highly influential time periods because they contribute to many features simultaneously.
Inspired by Kim \etal~\cite{kim2021towards}, we consider three design alternatives (Fig. \ref{fig:alternative}\cc{C}-\cc{E}) for highlighting prominent regions in the line chart. 
For \cc{C}, the bordered bounding box is accurate and clean, but not efficient at highlighting the overlapped area.
For \cc{D}, the translucent full-height box highlights the overlap well, but is visually crowded.
In accordance with our clinician collaborators, we finally chose the last design \cc{E} which combines the advantages of the other two designs.


\subsection{Timeline View}
\label{sec:view_timeline}

The \textit{Timeline View} (Fig. \ref{fig:teaser}\cc{C}) provides an overview of the target patient's health records (\ref{r:record-summary}).
This view is the starting point for clinicians who use the forward analysis workflow (\ref{f:forward}). In the meantime, it is also an indispensable part of the backward analysis workflow (\ref{f:backward}), when a clinician desires to understand more contextual information about the patient.
Through this view, clinicians can move additional medical records into the \textit{Temporal View} for comparative analysis. 


We use a matrix-based visualization~\cite{weng2021towards} to show a summary of the target patient's medical events from different sources (lab tests, vital signs, and chart events) (Fig. \ref{fig:teaser}\cc{C}).
The horizontal timeline is divided into predefined, equal time intervals (\eg, 1h, 4h, and 8h).
Each cell contains the two pieces of information our clinician collaborators deemed most vital: (1) the background color encodes the number of events, with darker blue representing more events; and (2) the width of the inner box encodes the proportion of events containing out-of-reference-range values.
For example,  
\begin{tcolorbox}[boxrule=1mm,colframe=lightblue, height=3mm, width=3mm]
\end{tcolorbox} 
indicates that very few events occurred during this period and that most of them were normal, while 
\begin{tcolorbox}[boxrule=0.5mm,colframe=darkblue,height=3mm, width=3mm]
\end{tcolorbox}
has the opposite meaning, and may call for an in-depth inspection.
A similar design was used in Voila~\cite{cao2017voila} to visualize the number of anomalous events of a region on a map.  

Observing interesting cells in a particular row (\eg, lab tests), clinicians can brush to select them, and click the ``Go Temporal View'' button to visualize all records from different items (\eg, lab test items such as ALT, Glucose, and Lactate) in the \textit{Temporal View} for a detailed investigation and comparative analysis. 

\subsection{Interaction}
In addition to the basic interactions introduced above, \revise{we propose two novel interactions}{\system{} offers two additional interactions}, \textbf{linking} and \textbf{marking}, to facilitate better visual associations between features and their corresponding records. 

\textbf{Visually associating features and medical records (\ref{r:bridge-association}).}
Understanding connections between the feature elements (\ie, rows in the feature list), and medical record elements (\eg, temporal records in line charts and static information listed in the patient's profile) is not easy. Clinicians may need to scroll through a long list of features and compare names one-by-one. 
To make this easier, we propose the following novel and intuitive strategy. 
First, we use small colored bars to indicate the data source (\eg, lab tests and vital signs) for both feature elements (on the right border) and medical record elements (on the left border). 
Then we draw curves to connect the associated feature elements with medical record elements (Fig. \ref{fig:teaser}\cc{\small{E2}}).
These curves are dynamically updated when users scroll down or join additional time-series records into the \textit{Temporal View}. 

\textbf{Marking on medical records.}
To support the forward analysis workflow, clinicians are allowed to mark interesting medical record elements with ``pins'' (Fig. \ref{fig:teaser}\cc{\small{E1}}). Associated feature elements are highlighted with a thicker bar. Clinicians can temporarily remove all other irrelevant features or feature groups by turning on the ``focus'' switch in the feature view's left-top corner.

\section{Evaluation}
In this section, we first introduce two case studies conducted with two clinicians (\textbf{P1}, \textbf{P5}) for evaluating whether \system~and our proposed workflows (\ref{f:forward}, \ref{f:backward}) can support clinical decision-making. 
All clinicians also participated in the pilot study and development process, and are therefore familiar with the system.

\subsection{Case Study I - Backward Analysis}

We worked with \textbf{P5}, who has 10 years of work experience, to explore and the model's predictions about a two-month-old infant admitted to the CICU.
The patient was predicted to be at high risk for various complications (L, C, A). 
The clinician was most interested in predicting cardiac complications (C), since they can lead to severe consequences.\revise{He expected that the explanations provided by \system~would help him better understand the patient's conditions and make postoperative care plans.}{}
He first selected a group of patients in the same age range (\ie, infants from 28 days to 12 months) to serve as references (\ref{r:feature-reference}, \ref{r:record-reference}). This group included 869 patients (Fig.~\ref{fig:teaser}\nnbox{0}) of which 550 were healthy.

\textbf{Exploring feature hierarchy (\ref{r:feature-group}).}
The clinician first glanced at the patient's profile and noticed that two features \nnbox{1}, surgery time and CPB (cardiopulmonary bypass) time, were much higher than usual. Keeping this in mind, he started exploring the feature view to check the features' contributions to the predicted complications.
In the top level of the feature hierarchy \nnbox{2}, he noticed that the contribution bar of the ``in-surgery'' feature group was much longer than that of the ``pre-surgery'' feature group, which means that the model mainly used information collected during surgery to make the prediction.
The clinician then expanded the feature hierarchy and zoomed into a lower level to inspect the detailed explanations. 
Through sorting and filtering, he settled on a configuration where only the top 5 features or groups with the highest contributions were displayed in the list (\ref{r:feature-group}, \nnbox{3}). ``\textit{I like this control function and it helps me narrow down to a more focused display with only a few most important features}'', \mbox{he commented.}

\textbf{Understanding feature contributions (\ref{r:feature-reference}).}
He then noticed that the CPB time and the surgery time were the top 2 most important features whose values were both out of distribution \nnbox{3}. He then commented ``\textit{This is exactly what I expected. Great to have a confirmation from the model about my previous suspicion}''. 
He further wondered ``\textit{What would happen if their values go back to normal?}''.
We reminded him of the what-if analysis function (\ref{r:feature-explore}).
Using this function, he found no noticeable change in the prediction results for both features. 
However, he noticed that reducing the surgical time to the normal range decreased its contribution significantly.
He thought ``\textit{The exceptionally long surgical time makes this feature positively contribute a lot to the model prediction, but other factors are still playing important roles because the prediction result does not change after what-if}''.


The clinician then moved on to the two other features of interest -- Oxygen Saturation and Lactate -- as they are critical indicators of a patient's condition.
Zooming into the most fine-grained feature level \nnbox{4}, he discovered that the contributions of these two features mostly came from the Mean features\footnote{Other features such as SD and Trend were filtered out due to their insignificant contributions.} which were either exceptionally low (Oxygen) or high (Lactate).
He suspected such abnormal values should have considerable impacts on the model prediction and confirmed this suspicion after what-if analysis \nnbox{4}.
He then showed further curiosity about the details of these abnormal values and commented ``\textit{I have to figure out when and why the lactate/oxygen saturation started to accumulate/drop. This is important for me to understand which catalysts, such as a patient's pre-existing condition or a surgeon's mistake, cause the results.}''.
So he selected the corresponding features to review them in the \textit{Temporal View} (\ref{r:bridge-association}). 

\textbf{Inspecting features' influential records (\ref{r:record-reference}).}
After the temporal view was displayed, he immediately observed that the Lactate level \nnbox{5} was normal at the beginning of surgery, but started to increase after 2:00 PM and eventually went above the reference range after 3:00 PM. 
However, the Oxygen Saturation \nnbox{6} was below the reference range during almost the entire surgical period (all the red downward-facing arrows).
He commented ``\textit{I am so impressed by the smooth interaction and intuitive visualization design to guide me here. I think this patient might have cyanotic congenital heart disease, which could be the root cause for the hypoxemia and the lactate accumulation.}''.
He then decided to continue exploring the direct reason for the Lactate accumulation.
He hypothesized that such accumulation was directly caused by the CPB process\footnote{CPB is a technique that temporarily takes over the function of the heart and lungs during surgery, maintaining the circulation of blood and oxygen.}. To confirm this, he referred to the timeline view and selected the vital signs during the surgery \mbox{as references \nnbox{7} (\ref{r:record-summary})}. 

Taking a close look at the Pulse records, he noticed that the Pulse dropped to a very low level (50 BMP) at 3:00 PM and returned to normal at 5:00 PM \nnbox{8}.
He confirmed this was the CPB period and explained ``\textit{During this time, the functions of the patient's heart and lungs were taken over by the CPB pump. That's why the patient's pulse looks abnormal.}''
Comparing this period with the Lactate curve, he then rejected his earlier hypothesis, because the lactate had already reached a high level at 3:11 PM and in that case, the accumulation would have started earlier. 
Another interesting pattern -- a sudden drop of Pulse around 2:30 PM \nnbox{9} -- caught his attention. He thought ``\textit{This was a rescue conducted at that time and is likely to be the key reason accounting for lactate accumulation}''.

Noticing the sudden drop in Pulse, he was curious about whether the model ``captured'' this information while making predictions (\ref{r:bridge-highlight}).
He then clicked the ``explain'' button to toggle the influential segments from the model's point of view \nnbox{8}.
He noticed that most of the orange (influential) areas covered the CPB period.
This fine-grained explanation is slightly different from his expectation -- from his perspective, the model should also pay attention to the former sudden drop in Pulse.
But in general, he agreed that the prediction was based on the most potentially critical medical records, and was trustworthy.

\textbf{Summary.} 
Through this exploration, \textbf{P5} was able to understand the most important features that led to the prediction, and to explore some interesting features and their corresponding records in depth.  
He decided to pay more attention to this patient, and considered using proactive treatments to avoid the situation getting worse \mbox{in postoperative care.}


\subsection{Case Study II - Forward Analysis}

\begin{figure}[!t]
    \centering
    \includegraphics[width=1\linewidth]{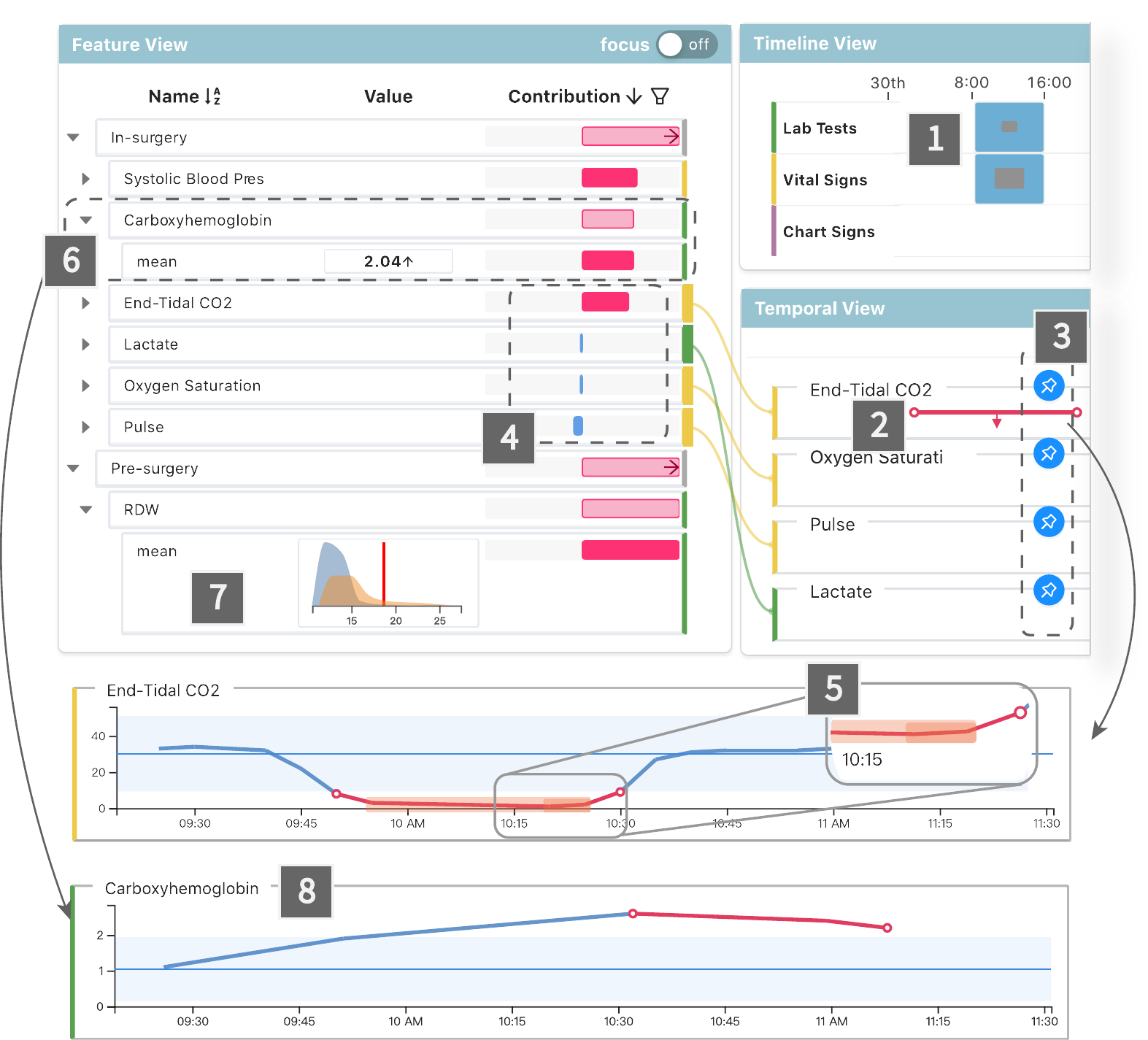}
    \vspace{-0.2in}
    \caption{Case Study II -- a use case involving understanding a prediction of high-risk lung-related complications following a \textbf{forward analysis workflow}. 
    The clinician gained an overview of the patient's in-admission records from the timeline view~\nnboxx{1}. 
 She then inspected the record details~\nnboxx{2}, marked interesting items~\nnboxx{3}, and formed the hypothesis. 
    She verified the hypothesis with feature contributions~\nnboxx{4} and influential time segments of the marked items~\nnboxx{5}, which were mostly within expectations. 
    Finally, she explored features with unexpectedly high contributions \nnboxx{6}-\nnboxx{8}, which helped her refine her judgements. 
    }
    \label{fig:case-2}
    \vspace{-1.7em}
\end{figure}

We worked with \textbf{P1} -- who has 20 years of  experience in this field -- to understand a prediction of high-risk lung-related complications made by the machine learning model.

\textbf{Gaining an overview of patient information (\ref{r:record-summary}).} 
The clinician started by checking the patient profile view. She thought everything (\eg, surgical time and CPB time) was normal except for the patient's age (11 months), which was young for a VSD repair surgery.
Then she looked at the timeline view and found the period during surgery (Fig. \ref{fig:case-2}~\nnboxx{1}).
In the row of lab tests, she noticed that most in-surgery test results were in normal ranges, indicated by the small grey inner rectangle. 
At the same time, vital signs had a slightly higher proportion of abnormal records. After the initial exploration, she found no solid evidence to indicate complication risks.

\textbf{Inspecting record details (\ref{r:record-reference}).}
She then checked the detailed lab tests and vital signs~\nnboxx{2}.
She commented ``\textit{I don't find any big things. The three important indicators, Oxygen Saturation, Pulse, and Lactate, all look clean with no anomalous segments.}''.
She also noticed End-Tidal CO2 was below the reference range for a long period. \revise{}{Nevertheless, she hypothesized that the patient was not likely to have complications, which contradicted the model prediction. So she planned to refer to model explanations to figure out whether there were factors she had overlooked.}
She marked all four items~\nnboxx{3} and continued to check the explanations in the feature view. 

\textbf{Comparing feature contributions with expectations.}
By tracing the links to the feature list (\ref{r:bridge-association}), she noticed that the feature group related to End-Tidal CO2 had a high positive contribution to the high-risk prediction~\nnboxx{4}. 
In contrast, features related to the other three items had slight negative contributions. 
She praised \textit{``The explanation algorithm looks amazing. This actually matches what I expected. Now I am curious to see what the influential periods the model thinks to be''}.
She clicked the ``explain'' button for help and then obtained the orange-highlighted area~\nnboxx{5} which she thought was caused by CPB. 
The overlapped area with a deeper color also caught her attention, because multiple sub-features identified this area.
She then said \textit{``This is the critical changing point, but I might need more contextual information to test my thoughts''.}

She also noticed that Systolic Blood Pressure, Carboxyhemoglobin (COHb), and pre-surgery Red Cell Distribution Width (RDW) had the highest contributions. Among these, she noticed that the mean value of COHb~\nnboxx{6} and RDW~\nnboxx{7} was higher than the reference range. 
She commented \textit{``This is beyond my expectation. I know COHb is used to detect carbon monoxide (CO) toxicosis, but I never use this to judge whether a patient will develop complications''.}
Through further inspection~\nnboxx{8}, she found that the COHb level was the highest right after the abnormal segment of End-Tidal CO2. 
She thought \textit{``This might be unnoticeable factor in identifying the complications and I want to do further study with my team about it''.}
As for the high RDW level, she realized that it might indicate that the patient suffered from iron-deficiency anemia, making them vulnerable to (lung) infections. This lab test does not tend to draw much attention from cardiac surgeons, so she had missed it earlier. 

\textbf{Summary.} After the exploration, \textbf{P1} agreed that the patient was likely to have lung-related complications and decided to pay more attention to her. She was also curious about how COHb can be used to identify complications and considered studying it further.

\section{Discussion}

These case studies suggest that \system~is helpful to clinicians and can support them in their decision-making. 
\revise{The participants (\textbf{P1} and \textbf{P5}) were both able to use the system to verify their hypotheses and confirm their judgements following their own preferred workflows. }{}In addition to the case studies, we conducted semi-structured interviews with \textbf{P4} and \textbf{P6} by showing them the case study results and encouraging them to freely explore the system to collect additional feedback. We report and discuss feedback from all 4 clinicians as follows.

\subsection{Design Implications}
Feedback from these 4 clinicians led us to a set of important design considerations for all such projects, which we summarize as follows:

\textbf{Applications of \system{}.}
All 4 participants generally commended the usefulness of \system~in supporting diagnoses and expected to use the system to improve their daily workflow.  
\textbf{P1} expected to use the system to \textbf{make more accurate decisions}. She said \textit{``Everyone sometimes may fall into blind spots. This tool can actually help me reduce the risk of making mistakes''}. 
\textbf{P4} expected to use the system to \textbf{communicate better} with other clinicians. 
He commented that \textit{``A surgery involves collaborations between teams, $\dots$, people see data from different angles which might be biased somehow, $\dots$, I would trust \system~and believe it can greatly facilitate the communication between teams''}.
Both \textbf{P5} and \textbf{P6} suggested using \system~to \textbf{help junior doctors} to make more accurate diagnoses.

\textbf{Reference-value-based explanations.}
The reference values are vital in facilitating prediction interpretations for clinicians. Explanations like ``the $Pulse.Mean$, whose value is below the reference range, has a high contribution to patient's cardiac complication'' are easier for clinicians to understand and accept than purely reporting the contribution scores as confirmed by \textbf{P5}.

\textbf{Feature hierarchy design.}
The hierarchical display of features was praised by all participants as it helps them avoid unnecessary details during exploration. 
\revise{P1 suggested that we refer to commonly used medical scales to organize the feature hierarchy (\eg, grouping the surgery name and the patient's conditions as ``surgical risks''), so that they would be more familiar to clinicians.}{Currently, there is no standard for designing the hierarchy of all healthcare features. However, ideas can be borrowed from the clinical forms used for communications between clinicians as suggested by \textbf{P1}.} 

\textbf{Providing explanations with context.}
\revise{The fourth takeaway is to align explanations with context using various interactions.}{}As demonstrated by the case studies, contextual information helps clinicians to understand explanations. Those in our study appreciated how the various visualization and interaction techniques in the system facilitated visual association between explanations and context. \textit{``With the links, I can easily get connections between the features with their corresponding results,''}, as \textbf{P4} said. Also, \textbf{P1} suggested that ``marking'' is a very convenient interaction for checking explanations at will. 



\subsection{Limitations and Future Work}
\revise{We identify two limitations in our work.}{We introduce the limitations of our current work and future plans.}

\textbf{Feature interpretability.}
Our system only focuses on explaining predictions made from interpretable features (\ie, features that have clear meanings and are extracted from a series of relevant health records). 
When the feature itself is hard for humans to understand (\eg, features built from representation learning methods), the connections between features and health records can be very complex.
In this case, the system will be less effective.
An advanced method for tracing and storing such complex connections would be a good addition and remains to be explored.

\revise{}{
\textbf{Potential cognitive biases.}
Wang \etal's work~\cite{wang2019designing} suggests that a backward-oriented reasoning process (\ie, first acquiring the diagnostic predictions, and then looking for supporting evidence) may lead to confirmation bias. Potential effects of cognitive biases on clinicians' decision-making when following different analysis workflows, and how our visualization designs may alleviate potential risks, have not been fully evaluated in this work. We plan to study this further by assessing the precision of clinicians' decisions when using \system{}.}

\revise{}{
\textbf{Quality of EHR data.}
The poor quality of EHR data (\eg, missing data) is a challenge to EHR data analysis in general. During the \system's development process, we also found many ``False Positive'' patterns caused by misrecorded data items (\eg, a seeming cardiac arrest pattern was traced back to a faulty sensor). Currently, clinicians' prior knowledge is required to detect these data defects. In the future, we plan to investigate anomaly detection and visualization solutions to detect and encode any missing information in order to raise clinicians’ awareness of missing data.}

\revise{}{\textbf{Precision of reference values.}
To improve the usability and precision of the dynamic reference value selection method, we plan to make the following extensions. First, we will automatically recommend relevant cohorts to clinicians for obtaining reference values. Second, we will derive time-varying reference values for temporal records (\eg, Pulse), which are more applicable to surgical scenarios that are composed of multiple stages. Third, we will conduct experiments to understand the stability of the reference values (\ie, how will the reference values change as the cohort changes over time?).} 

\textbf{Visual scalability.}
Scalability issues occur in the temporal view when analyzing a signal with a large number of records.
In addition, as the number of test items (rows) increases, finding interesting ones becomes less efficient, as more scrolling is required. 
In the future, we plan to scale up our approach by 1) segmenting long signals in different scales and 2) using searching and filtering techniques to facilitate the exploration of a vast number of complex signals. 

\revise{}{\textbf{Generalizability to other healthcare models.}
\system{} can be generalized to work on other prediction problems (\eg, mortality predictions) and other ML models using the PIC dataset. However, adaptations (\eg, formal descriptions of the entities and generated features) must be made to use \system{} with other EHR datasets (\eg, MIMIC-III~\cite{johnson2016mimic}), which is required by the feature extraction process (introduced in Sec.~\ref{sec:modeling:feature_abstraction}). In the future, we plan to improve generalizability by defining system inputs according to the Fast Healthcare Interoperability Resources (FHIR) standard~\cite{FHIR}, a general EHR data format.} 


\section{Conclusion}
In this work, we identified three key challenges limiting the use of ML in clinical settings, including clinicians’ unfamiliarity with ML features, lack of contextual information, and the need for cohort-level evidence.
We then introduced \system~ -- a visual analytics system designed according to the requirements identified in a pilot study -- to support clinicians using ML to make decisions with both forward and backward analysis workflows. 
We conducted two case studies and expert interviews with 4 clinicians. Their positive feedback and in-depth insights demonstrate the usefulness and effectiveness of the system.
In particular, it reveals that visually associating model explanations with patients’ situational records can help clinicians better interpret model predictions and use them to make clinical decisions. 
\revise{In the future, we plan to address the aforementioned issues and consider extending the system to support more data types, especially unstructured data such as clinical notes and medical images.}{}

\acknowledgments{We would like to thank Jianchuan Qi, Jie Jin, Lianglong Ma, Shanshan Shi, Xiuning Wu, and Xucong Shi from the Children’s Hospital of Zhejiang University School of Medicine for their insightful feedback during the design process. We thank Arash Akhgari for his efforts in making the figures and feedback on the system interface designs. We also thank the anonymous reviewers for their valuable comments.

This research was supported in part by Hong Kong Theme-based Research Scheme grant T41-709/17N and a grant from MSRA.}

\clearpage

\bibliographystyle{abbrv}
\balance
\bibliography{bibliography}

\end{document}